# Realization of low-energy Dirac fermions in $(Ir_{1-x}Pt_x)Te_2$ superconductors

B.-B. Fu,[1,2,#] C.-J. Yi,[1,2,#] Z. J. Wang,[3,#] M. Yang,[1,2] B. Q. Lv,[1,2] X. Gao,[1,2] M. Li,[4,5] Y.-B. Huang,[4] C. Fang,[1] H.-M. Weng,[1,6] Y.-G. Shi,[1,*] T. Qian,[1,6,*] and H. Ding[1,2,6,*]

[1] *Beijing National Laboratory for Condensed Matter Physics and Institute of Physics, Chinese Academy of Sciences, Beijing 100190, China*

[2] *University of Chinese Academy of Sciences, Beijing 100049, China*

[3] *Department of Physics, Princeton University, Princeton, NJ 08544, USA*

[4] *Shanghai Synchrotron Radiation Facility, Shanghai Institute of Applied Physics, Chinese Academy of Sciences, Shanghai 201204, China*

[5] *Department of Physics and Beijing Key Laboratory of Opto-electronic Functional Materials & Micro-nano Devices, Renmin University of China, Beijing 100872, China*

[6] *Collaborative Innovation Centre of Quantum Matter, Beijing, China*

[#] These authors contributed to this work equally.

[*] Corresponding authors: ygshi@iphy.ac.cn, tqian@iphy.ac.cn, dingh@iphy.ac.cn

**Abstract**

Superconducting and topological states are two distinct quantum states of matter. Superconductors with topologically nontrivial electronic states provide a platform to study the interplay between them. Recent study revealed that a bulk superconductor PdTe$_2$ is a topological Dirac semimetal with type-II Dirac fermions. However, the Dirac fermions in PdTe$_2$ have no contribution to the superconducting pair because the Dirac points reside far below the Fermi level ($E_F$). Here, we show that in the isostructural compounds (Ir$_{1-x}$Pt$_x$)Te$_2$, the type-II Dirac points can be adjusted to $E_F$ by element substitution at $x \sim 0.1$, for which the bulk superconductivity appears near 2 K. The (Ir$_{1-x}$Pt$_x$)Te$_2$ superconductor with the Dirac points at $E_F$ paves the way for studying the relevant exotic physical phenomena such as topological superconductivity in Dirac semimetals.

The interplay between topological states and superconductivity is an important issue in condensed matter physics [1,2]. One sequence of the interplay is Majorana bound states or Majorana fermions at the edge of topological superconductor, which can be used for topological quantum computations. Superconducting topological materials are promising candidates to realize topological superconductivity. Superconductivity has been observed in topological insulators such as Cu-intercalated $Bi_2Se_3$ [3]. Very recently, it has been discovered that the iron-based superconductor $FeTe_{0.55}Se_{0.45}$ with a superconducting transition temperature $T_c$ = 14.5 K has a nontrivial topology with spin-helical topological surface states (TSS). Furthermore, the superconducting gap of those TSS has also been observed, paving a new way to realize Majorana fermions at higher temperature [4, 5].

As with topological insulators, topological semimetals also have a nontrivial topology [6-10], where superconductivity could be realized. Superconductivity can be induced by a point contact [11,12] or under high pressure [13] in the topological Dirac semimetal $Cd_3As_2$. The point contact spectra of $Cd_3As_2$ show a zero-bias conductance peak, suggesting unconventional superconductivity around the point contact region [11,12]. Distinct from type-I Dirac semimetal like $Cd_3As_2$, type-II Dirac semimetal phase possesses over-tilted Dirac cones sharing similar mechanism with type-II Weyl semimetal phase [14]. Recent studies have revealed that the type-II Dirac semimetal state can be realized in several transition metal dichalcogenide compounds $PtTe_2$ [15], $PdTe_2$ [16], and $PtSe_2$ [17] with the $CdI_2$-type 1T structure. Among them, $PdTe_2$ has the bulk superconductivity with $T_c$ = 1.7 K at atmospheric pressure. However, as the Dirac points reside at 0.5 eV below $E_F$ in $PdTe_2$, the superconductivity may have nothing to do with the Dirac points.

To study the interplay between Dirac fermions and superconductivity, it is necessary to adjust the Dirac point to $E_F$ while retaining the superconductivity. To lower $E_F$ to the Dirac point, one common route is to reduce the number of valence electrons by element substitution. The number of valence electrons of $IrTe_2$ is one less than that of $PdTe_2$. $IrTe_2$ has a structural transition at 270 K [18] and its high-temperature phase is isostructural to the 1T-structure $PdTe_2$ (Fig. 1a). The first-principles calculations indicate that the 1T-structure $IrTe_2$ has a similar electronic structure, in which the band crossing along ΓA leads to a type-II Dirac point lies at

0.18 eV above $E_F$ (Fig. 1e,f). Unfortunately, no superconductivity has been found in the pure 1T phase.

By substituting Pt or Pd with Ir, the structural transition is dramatically suppressed, which is accompanied with the appearance of bulk superconductivity [18]. In the phase diagram of $(Ir_{1-x}Pt_x)Te_2$, $T_c$ reaches a maximum 3 K at $x = 0.05$ and gradually reduces with further increasing the Pt content [18]. We have synthesized a series of $(Ir_{1-x}Pt_x)Te_2$ samples with $x$ up to 0.5. The EDX measurements confirm that the chemical compositions are consistent with the corresponding nominal ones (Fig. 2b). Our measurements of resistivity as a function of temperature show that $T_c$ is about 2 K for $x = 0.1$ and the superconductivity does not appear at temperature above 0.3 K for $x = 0.25$ (Fig. 2a).

It is expected that the Dirac points should pass through $E_F$ with the Pt substitution, assuming that there are no essential changes in the electronic structure. To determine the composition that the Dirac point passes through $E_F$, we have investigated the electronic structures of the samples with different doping levels using angle-resolved photoemission spectroscopy (ARPES). Apart from EDX measurements, the results of core level measurements further help us confirming the various compositions qualitatively, as shown in Fig. 2c. The intensity of Ir 4f peaks decreases and Pt 4f peaks increases monotonically, with the normalization of Te $4d_{5/2}$ peaks, when the content of Pt raised.

In Fig. 3, we display the ARPES results of the $x = 0.5$ sample. The calculations indicate that the Dirac point is located on the high-symmetry line ΓA along the $k_z$ direction. As the ARPES experiments were preformed on the (001) cleavage surface, we measured the band structures in the $k_x$-$k_y$ planes at different $k_z$ locations by varying the photon energy ($hv$) of incident lights. Figure 3a shows the band dispersions along the ΓK direction taken with different $hv$'s. In Fig. 3a(i), we observe one electron band near $E_F$ and one hole band below -0.2 eV around $k_x = 0$ at $hv = 20$ eV. With increasing $hv$, both the electron band and hole band gradually sink down with different rates and touch with each other (Fig. 3a(iii)), and then separate, forming the tilted type-II Dirac bands. as seen in Figs. 3a(i)-3a(v). In Fig. 3b, we plot the extracted dispersions of the two bands measured at different photon energies. The two bands are separated except that they touch each other at $hv$ = 22 eV. Figure 3c exhibits the band dispersions in $k_x$-$k_y$ plane along the S-D-T direction, denoted in Fig. 1b. Figure 3d shows the stacking ARPES intensity maps at several constant energies. There is a circular-like Fermi surface around the Brillouin zone center at $E_F$. The Fermi surface shrinks into a point at 0.27 eV below $E_F$ and then grows into a circle again with further lowering the energy. We extract the band dispersions along $k_z$, indicated as green dashed lines in Fig. 3b. The velocities of the two bands along $k_z$ have the same sign. These results confirm that the touching point observed at $hv$ = 22 eV is a type-II Dirac point, which is nearly isotropic in the $k_x$-$k_y$ plane but strongly tilted along $k_z$. The Dirac point of the $x$ = 0.5 sample lies at 0.27 eV below $E_F$, which is closer to $E_F$ as compared with that in PtTe$_2$, where the Dirac point lies at 0.97 eV below $E_F$ [15].

Next we focus on the evolution of band structure with the Pt substitution. For all the samples with different Pt contents, the measured band structures are almost consistent except for the energy positions relative to $E_F$. Figure 4a-d shows the band dispersions along the ΓK direction taken at $hv$ = 22 eV of the (Ir$_{1-x}$Pt$_x$)Te$_2$ samples with $x$ = 0.05, 0.1, 0.25, and 0.5, respectively. The results clearly show that the Dirac point monotonically shifts upwards with the decrease of Pt content. From the extracted Dirac bands of the measured samples with different Pt contents in Fig. 4g, we determine that the Dirac point passes through $E_F$ at $x$ ~ 0.1, for which the ARPES intensity map at $E_F$ exhibits a point-like Fermi surface at the Brillouin zone center (Fig. 4f). Figure 4h shows that the energy position of the Dirac point does not shift linearly with the doping level. The energy shift of the Dirac point becomes faster at higher Pt contents. The extrapolation based on the trend is consistent with the previous result of PtTe$_2$ [15]. We mention that if rigidly shifting the calculated bands of IrTe$_2$, the Dirac point would pass through $E_F$ around $x$ ~ 0.25. This value is higher than that determined from our experimental results. This may be due to that the band calculations slightly overestimate the energy position of the Dirac point of IrTe$_2$. The previous study in PtTe$_2$ shows similar inconsistency between calculation and experiment [15], with a discrepancy of about 0.1eV.

From our experimental results, we determine that the Dirac point lies at $E_F$ in the $x$ = 0.1 sample with bulk superconductivity of $T_c$ = 2 K. This temperature is achievable in many experimental techniques, making the study of physical properties of the Dirac fermions in the superconducting state possible. Actually, the substitution of Pt in 1T-strucuture IrTe2 has two benefits: First, the substantial electron-doping lifts up the $E_F$, leading the type-II Dirac points more close to $E_F$. Secondly, it brings the bulk superconductivity to the system. In the region of $0.05 \leq x \leq 0.1$, both the superconductivity and the Dirac points at $E_F$ have been confirmed in our transport and APRES experiments, respectively. The theoretical studies have proposed that the topological superconductivity can be realized in superconducting type-I Dirac semimetal [19,20]. In contrast, there is no theoretical study on the superconducting properties in type-II Dirac semimetal. Our clear evidences suggest that the Pt-doping IrTe2 is an intriguing platform to further study the potential unconventional superconductivity in type-II Dirac semimetals. The relevant theoretical study is also called for in the near future.

*Note added*: During the preparation of the manuscript, we became aware of an arXiv paper [21], which reports similar ARPES results.

# Methods

**Sample synthesis and characterization.** Series of Pt doped Ir1-xPtxTe2 single crystals were synthesized by Te flux method in two steps. Firstly, Ir1-xPtx alloy (x=0.05, 0.1, 0.25, 0.5) was melted in an arc furnace under high-purity argon atmosphere. The alloy was melted several times for homogeneity. Secondly, the alloy and Te granules were put in an alumina crucible in a molar ratio of 1:8 and sealed in quartz tube under vacuum. Then the tube was heated to 1373K over 10 hours and dwelt for 20 hours. After that the tube was slowly cooled to 1073K in a rate of 2K/h and then separated by centrifuging. The chemical composition of single crystals of each doped concentration was analyzed by energy-dispersive x-ray spectroscopy (EDX) in a Hitachi S-4800 at an accelerating voltage of 15 kV. We measured several single crystals for each doped concentration and several areas in each crystal to confirm the homogeneity of the single crystals. Longitudinal resistivity was measured in an Oxford He-3 system in a configuration that four platinum wires were fixed on a crystal by silver epoxy.

**Angle-resolved photoemission spectroscopy.** High resolution ARPES measurement were performed at the 'Dreamline' beamline of the Shanghai Synchrotron Radiation Facility (SSRF) with a Scienta Omicron DA30L analyser and the 13U beamline of the National Synchrotron Radiation Laboratory (NSRL) at Hefei with a Scienta Omicron R4000 analyser, with the energy and angular resolutions set at 15 meV and 0.2°, respectively. All the samples were cleaved *in- situ* in vacuum condition better than 5 × $10^{-11}$ Torr and measured between 15 ~ 20 K.

**Band structure calculation.** We performed the first-principles electronic structure calculations of the 1T-structure IrTe2 with the projector augmented wave (PAW) method as implemented in VASP package [22]. The generalize gradient approximation of Perdew-Burke-Ernzerhof type was used for the exchange-correlation functional [23]. The cutoff energy for wave-function expansion was 500 eV. The k-point sampling grid is 16×16×8 for the band structure calculations. The tetrahedron method for Brillouin-zone integrations was adopted to computer the density of states (DOS). A denser k-point mesh of 24×24×18 was used in the DOS calculations.

**Fig. 1. Crystal structure and calculated electronic structure of IrTe$_2$. a**, Crystal structure. Orange and bule balls represent Ir and Te atoms, respectively. Black dashed lines indicate the unit cells. **b**, Bulk Brillouin zone. Red dots along Γ-A direction (labeled as D) mark the momentum locations of the Dirac points. **c,d,** Sketch maps illustating type-I and type-II Dirac cones, respectively. **e,f**, Calculated bands along Γ-A and S-D-T, respectively.

**Fig. 2. Superconductivity and chemical compositions of (Ir$_{1-x}$Pt$_x$)Te$_2$. a**, Resistivity as a function of temperature for the samples with $x$ = 0.05, 0.1, and 0.25. **b**, Representative EDX spectrum of $x$ = 0.1 sample determining the percentages of Ir, Pt, and Te elements, which are consistent with the nominal composition Ir$_{0.9}$Pt$_{0.1}$Te$_2$. The two insets show the real composition and scanned map with EDX. **c**, Shallow core level spectra of the samples with $x$ = 0.05, 0.1, 0.25, and 0.5. The two insets show the evolution of Ir 4$f$ and Pt 4$f$ peaks, respectively, with the Pt substitution.

**Fig. 3. Electronic structure of the (Ir$_{1-x}$Pt$_x$)Te$_2$ sample with $x$ = 0.5. a**, Band dispersions along the Γ-K direction. (i)-(v) were measured at $hv$ = 20, 21, 22, 23, and 24 eV, respectively. **b**, Stacking plot of the extracted bands from the data in **a**, showing the evolution of band dispersions near the type-II Dirac point. **c**, Stacking plot of the ARPES intensity maps at several constant energies measured at $hv$ = 22 eV, showing the Dirac cone structure in the $k_x$-$k_y$ plane.

**Fig. 4. Evolution of the Dirac point with the Pt substitution. a-d**, Band dispersions along the T-D-T direction measured at $hv$ = 22 eV for the (Ir$_{1-x}$Pt$_x$)Te$_2$ samples with $x$ = 0.05, 0.1, 0.25, and 0.5, respectively. **e**, Second derivative intensity plot of the data in panel **b**, showing the Dirac point lying at $E_F$ for the $x$ = 0.1 sample. Dashed lines are guide to the eye indicating the linear Dirac bands. **f**, ARPES intensity maps at $E_F$ measured at $hv$ = 50 eV for the $x$ = 0.1 sample, showing the point-like Fermi surface at the Brillouin zone center. **g**, Extracted Dirac bands from the data in panels **a-d**. **h**, Energy position of the Dirac point as a function of the Pt content. The value of PtTe$_2$ is extracted from the previous result in Ref. 12.

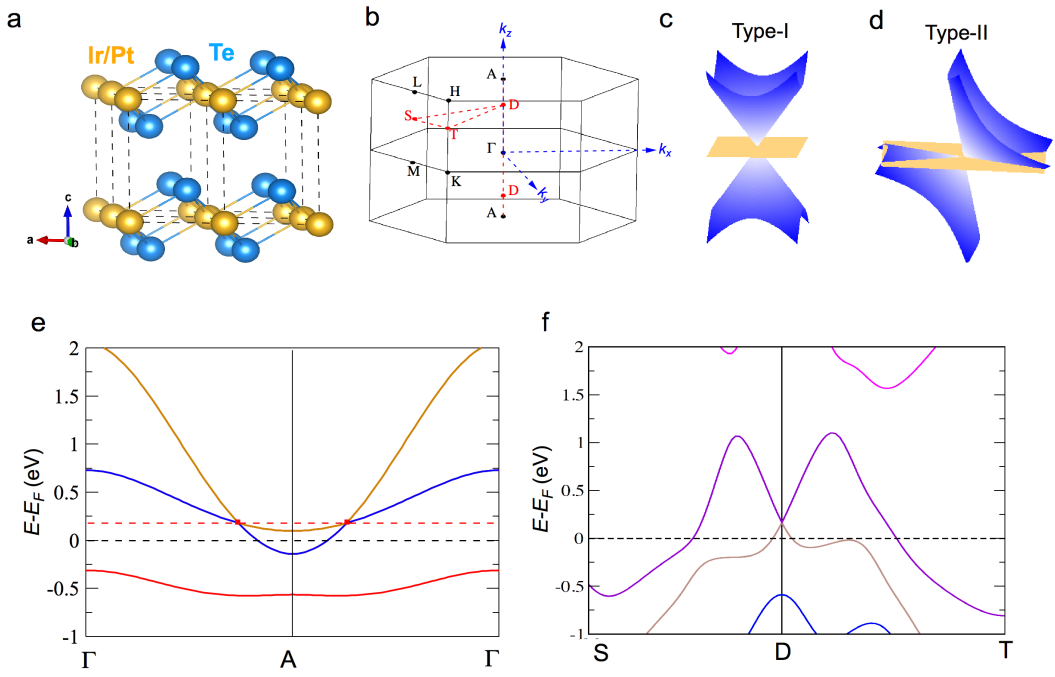

**Fig. 1. Crystal structure and calculated electronic structure of IrTe$_2$.**

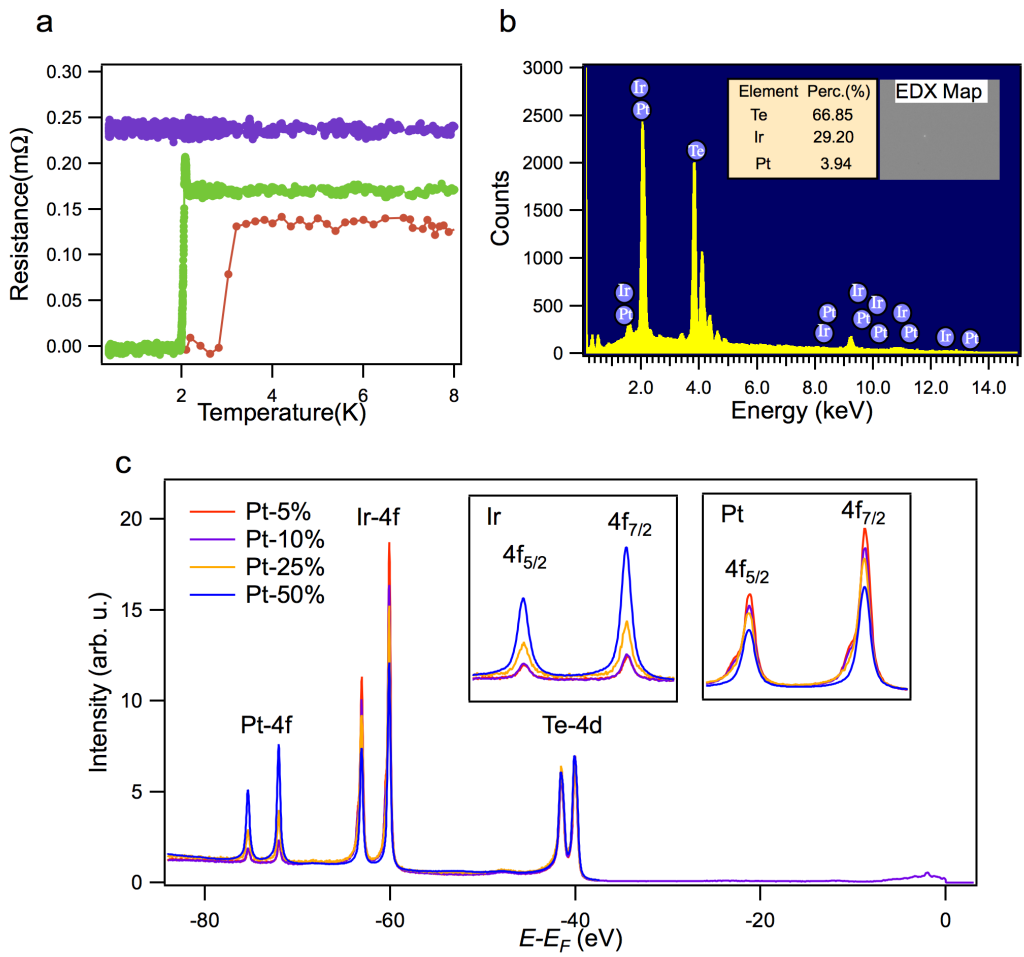

**Fig. 2. Superconductivity and chemical compositions of (Ir$_{1-x}$Pt$_x$)Te$_2$.**

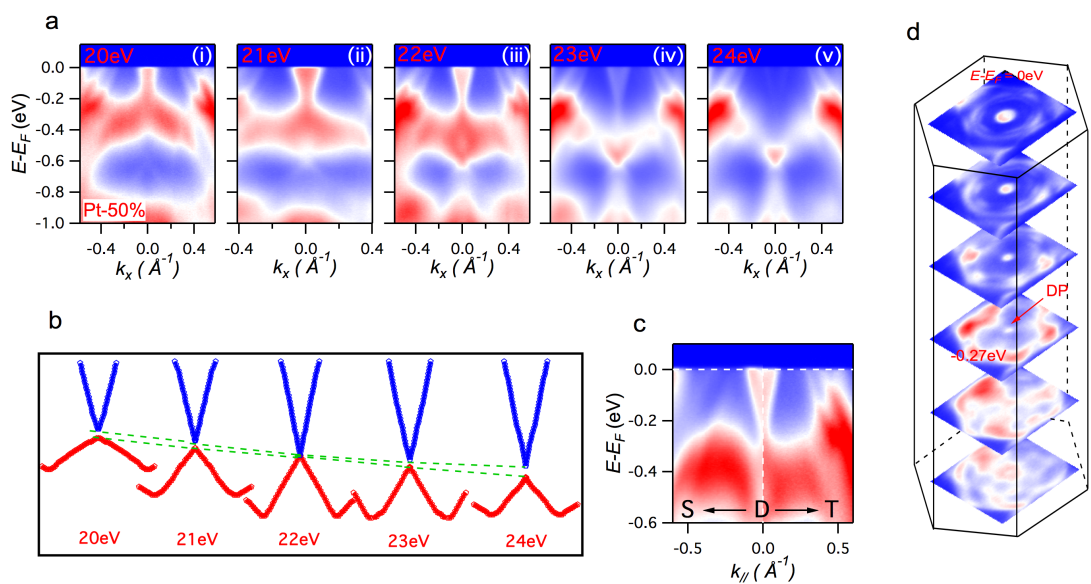

**Fig. 3. Electronic structure of the $(Ir_{1-x}Pt_x)Te_2$ sample with $x$ = 0.5.**

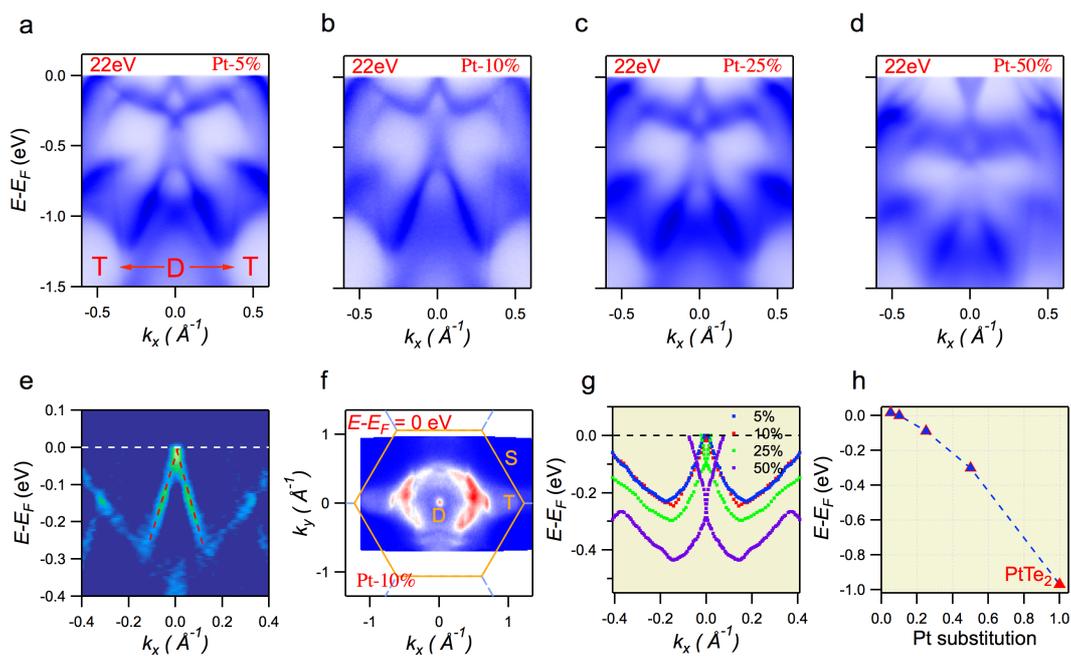

**Fig. 4. Evolution of the Dirac point with the Pt substitution.**